# Time Series Stock Price Forecasting Based on Genetic Algorithm (GA)-Long Short-Term Memory Network (LSTM) Optimization

Xinye Sha[1*]

[1]Columbia University, New York, 10027, USA
*Corresponding author email: xs2399@columbia.edu

**Abstract.** In this paper, a time series algorithm based on Genetic Algorithm (GA) and Long Short-Term Memory Network (LSTM) optimization is used to forecast stock prices effectively, taking into account the trend of the big data era. The data are first analyzed by descriptive statistics, and then the model is built and trained and tested on the dataset. After optimization and adjustment, the mean absolute error (MAE) of the model gradually decreases from 0.11 to 0.01 and tends to be stable, indicating that the model prediction effect is gradually close to the real value. The results on the test set show that the time series algorithm optimized based on Genetic Algorithm (GA)-Long Short-Term Memory Network (LSTM) is able to accurately predict the stock prices, and is highly consistent with the actual price trends and values, with strong generalization ability. The MAE on the test set is 2.41, the MSE is 9.84, the RMSE is 3.13, and the R2 is 0.87. This research result not only provides a novel stock price prediction method, but also provides a useful reference for financial market analysis using computer technology and big data.

## 1. Introduction

Stock price forecasting has been a much talked about research topic in the financial field and its background can be traced back to a long time ago. In the past, people mainly relied on factors such as fundamental analysis, technical analysis and market sentiment to predict stock price movements. However, with the development of computer technology and the advent of the big data era, time series algorithms play an increasingly important role in stock price forecasting.

Time series algorithm is a method of modelling and analysis using data arranged in chronological order [1]. In stock price forecasting, time series algorithms can help analysts and investors better understand the historical stock price movements, discover the patterns hidden behind the data, and make predictions of future price movements accordingly. Common time series algorithms include moving average [2], exponential smoothing [3], ARIMA model [4] (Autoregressive Integral Moving Average Model), LSTM (Long Short-Term Memory Network), and so on.

Time series algorithms can help to capture the patterns such as periodicity, trend and seasonality present in the data. By modelling and analyzing historical data, the algorithms can identify the patterns of stock price changes in different cycles, thus providing a reference for future trends [5]. The time series algorithm has strong flexibility and adaptability. As the market environment is constantly changing, traditional forecasting methods may not be able to adapt to the new situation. While time series algorithms can adjust and update the model in real time according to the latest data, keeping sensitive to market changes. In addition, time series algorithms are able to combine other factors for comprehensive analysis. In addition to historical stock price data, it can also take into account macroeconomic indicators, industry trends, the company's financial situation and other factors to improve forecast accuracy and reduce risk.

Time series algorithms play a crucial role in stock price forecasting, providing investors with a scientific and effective method to assist decision-making. In this paper, a time series algorithm based on Genetic Algorithm (GA) and Long Short-Term Memory Network (LSTM) optimization is used to forecast stock prices, which provides a new method for stock price forecasting.

## 2. Source of data sets

The dataset used in this paper is a private dataset. The data records the opening, high, low, and closing prices of the stock of a US-based multinational financial services company for the period from 2006/5/25 to 2021/10/11. For privacy reasons, we will refer to it as "the Global Fin Corp". The Global Fin Corp is headquartered in New York and the global operations are headquartered in Missouri. Worldwide, its primary business is the processing of payments between merchant banks and the issuing banks or credit unions of purchasers who make purchases using company-branded debit, credit and prepaid cards. Selected data are shown in table 1.

**Table 1.** The indicators and their meanings.

| Date | Open | High | Low | Close |
|---|---|---|---|---|
| 2006/5/25 | 3.748966684 | 4.283868673 | 3.739663824 | 4.279217243 |
| 2006/5/26 | 4.307126229 | 4.348057576 | 4.103397917 | 4.179679871 |
| 2006/5/30 | 4.183400226 | 4.184330423 | 3.986183781 | 4.093164444 |
| 2006/5/31 | 4.125722657 | 4.219679208 | 4.125722657 | 4.180608273 |
| 2006/6/1 | 4.179678186 | 4.474571934 | 4.176887151 | 4.419686317 |
| 2006/6/2 | 4.511782082 | 4.530387351 | 4.352706872 | 4.371312141 |
| 2006/6/5 | 4.376895226 | 4.581553703 | 4.372243796 | 4.572250843 |
| 2006/6/6 | 4.649463041 | 4.709930299 | 4.446665399 | 4.493178368 |
| 2006/6/7 | 4.495967881 | 4.502479703 | 4.348986117 | 4.428058624 |
| 2006/6/8 | 4.428061429 | 4.439224243 | 4.232705752 | 4.439224243 |
| 2006/6/9 | 4.539691402 | 4.539691402 | 4.435501776 | 4.444804192 |
| 2006/6/12 | 4.46527062 | 4.488526882 | 4.307125554 | 4.367592812 |
| 2006/6/13 | 4.297822402 | 4.404803063 | 4.214098888 | 4.232704163 |

The data were statistically summarized and mean, std, min, 25%, 50%, 75% and max were calculated for each data and the results are shown in Table 2.

**Table 2.** Data statistics.

| | Open | High | Low | Close |
|---|---|---|---|---|
| count | | 3872 | 3872 | 3872 |
| mean | | 104.896814 | 105.956054 | 103.769349 |
| std | | 106.245511 | 107.303589 | 105.050064 |
| min | | 3.748967 | 4.102467 | 3.739664 |
| 25% | | 22.347203 | 22.637997 | 22.034458 |
| 50% | | 70.810079 | 71.375896 | 70.224002 |
| 75% | | 147.688448 | 148.645373 | 146.822013 |
| max | | 392.65389 | 400.521479 | 389.747812 |

## 3. Method

*3.1. Genetic algorithm*

Genetic algorithm is an optimization algorithm inspired by the theory of biological evolution and aims to solve complex optimization problems by simulating natural selection and genetic mechanisms. Its principles are based on Darwin's theory of evolution, which considers the concept of survival of the fittest and the transmission of favorable genes in a population [6]. Genetic algorithm generates, evaluates, selects and mutates individuals through continuous iterations to finally find the optimal solution. The schematic diagram of genetic algorithm is shown in Figure 1.

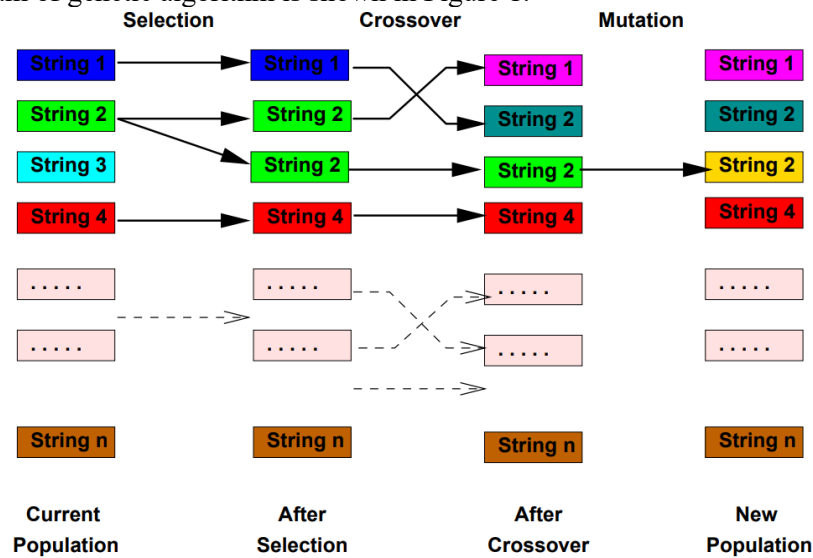

**Figure 1.** The schematic diagram of genetic algorithm.
（Photo credit : Original）

In genetic algorithms, a fitness function is defined to evaluate how well each individual performs in the problem space. The fitness function is usually designed according to the problem characteristics and is used to quantify the degree of superiority or inferiority of each individual. Secondly, initializing the population is the first step of the genetic algorithm, where a set of initial solutions are randomly generated as the population and the fitness value of each individual is calculated. In the selection phase, the individuals in the population are evaluated according to the fitness function and the better individuals are selected in a certain way to be the parents for reproduction. Common selection methods include roulette selection, tournament selection, etc., which ensures the principle of survival of the fittest. The selected sires will be involved in the breeding process. In the crossover (mating) stage, two individuals of the parent are selected and somehow exchange some of their genetic information to produce new offspring. Crossover operations can take various forms such as single point crossover, multipoint crossover, uniform crossover, etc., with the aim of retaining favorable genes in the parents and introducing new variants. This increases population diversity [7,8].

Next comes the mutation stage, in which mutation operations, i.e., randomly changing the values at certain loci, are performed on some of the progeny. The mutation operation helps to introduce new solution space to explore and improves the ability of the algorithm to jump out of the local optimum solution. It also helps to maintain population diversity.

Finally, in the replacement (updating) phase, the newly generated offspring are compared with the original population according to a certain strategy, the and the population is updated to retain the well-performing individuals. Often an elite or culling strategy is used to renew the population to ensure that the next generation contains better solutions. By repeating the above steps until a stopping condition is reached (e.g., the maximum number of iterations or convergence accuracy is reached), the genetic algorithm is able to search for a near-optimal solution in the problem space.

*3.2. Short- and long-term memory networks*
Long Short-Term Memory (LSTM) network is a deep learning model commonly used to process sequential data and is particularly adept at solving long sequential dependency problems. The principle of LSTM model is based on recurrent neural networks, which efficiently capture and store long term dependencies through the introduction of a gating mechanism [9]. The principle of the LSTM model is illustrated in Fig. 2.

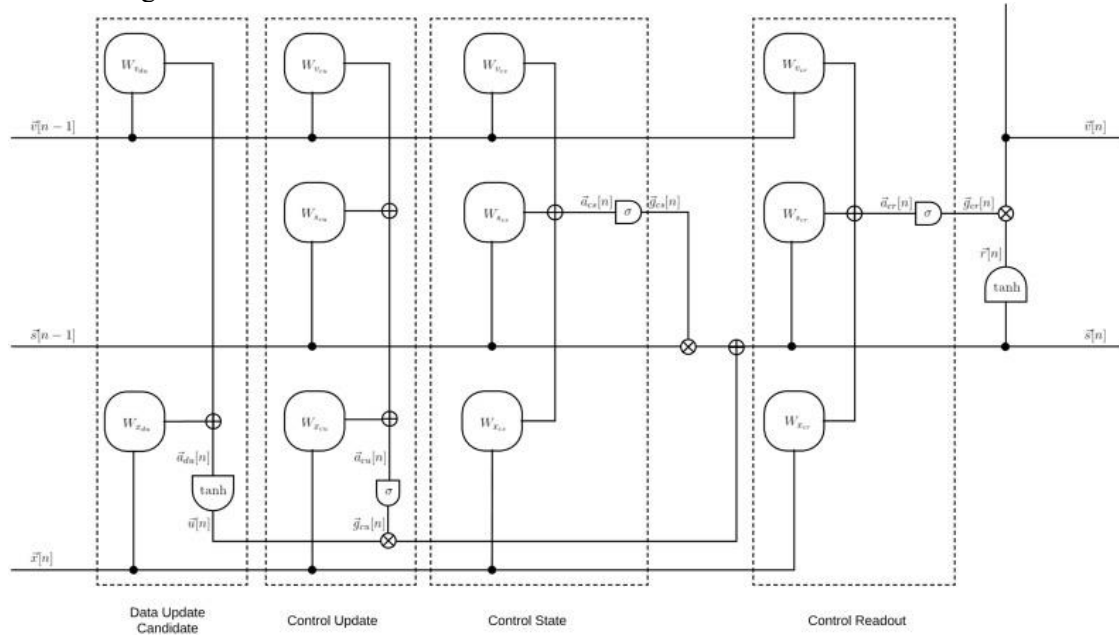

**Figure 2.** The principle of the LSTM model.
（Photo credit : Original）

The LSTM model contains three key components: input gate, forgetting gate and output gate. At each time step, the LSTM cell receives the input data, the hidden state of the previous moment and the cell state, and controls the flow and storage of information according to the gating mechanism.

Firstly, the input gate is responsible for deciding which information needs to be updated into the cell state. The input gate generates a value between 0 and 1 indicating the importance of each piece of information by means of a sigmoid activation function. Meanwhile, the tanh activation function generates a vector of candidate values for updating the cell state.

Next is the forgetting gate, which is used to decide which information needs to be forgotten from the cell state. The forgetting gate also generates a value between 0 and 1 through the sigmoid activation function, indicating the amount of information retained in each cell state. This effectively controls the effect of past information on the current state [10].

Finally, there is the output gate, which is responsible for determining the hidden state as well as the output value at the current moment. The output gate also generates a control output value ranging between 0 and 1 through the sigmoid function, while the tanh function is used to map the cell state to a new vector. Through the above three gating mechanisms, the LSTM model is able to achieve the modelling of long-term dependencies, and effectively solves the problems of gradient vanishing or explosion in traditional RNN models.

*3.3. Time Series Algorithm Based on Genetic Algorithm and Long Short-Term Memory Network (LSTM) Optimization*
Time series algorithms based on genetic algorithms and Long Short-Term Memory Network (LSTM) optimization combine the global search capability of genetic algorithms and the modelling capability of

LSTM models for sequence data to address the complexity and high-dimensional feature representation problems in time series forecasting.

Genetic algorithms are used to optimize the selection and tuning of hyperparameters for LSTM models to improve model performance and generalization. In genetic algorithms, an individual is usually represented as a combination of a set of hyperparameters, i.e., learning rate, number of hidden units, number of layers, etc. The performance of each individual is evaluated by a fitness function and the population is continuously evolved using operations such as selection, crossover, and mutation to find the optimal hyperparameter combination. When combining genetic algorithm and LSTM model, the LSTM model is first initialized or tuned using genetic algorithm to find a better initial solution. Then, during the training process, the hyperparameters can be continuously tuned to optimize the LSTM network structure according to the genetic algorithm, and the best individuals can be selected for updating in each iteration.

The LSTM model acts as a processor for time series data and can capture long-term dependencies and temporal features in the data. By training the LSTM model, complex patterns and trends in the data can be learned and predictions of future points in time can be achieved. LSTM has strong memory and nonlinear modelling capabilities, and performs well in time series prediction tasks.

In the prediction phase, an optimized LSTM model is trained to be used to make predictions about future time series. By inputting historical data sequences and combining them with the learned temporal patterns, LSTM can generate predictions for the corresponding time points. Such an approach combining the global search of the genetic algorithm and the time-series modelling capability of the LSTM model can improve the accuracy and robustness of time series prediction.

## 4. Experiments and Results

The dataset was divided and the data from 2016-2021 was selected as the training set and the data from 2021-2022 was selected as the test set as shown in Fig. 3. Epoch was set to 100 and run using Python 3.10 and the model was evaluated using MSE, MAE and MAPE.

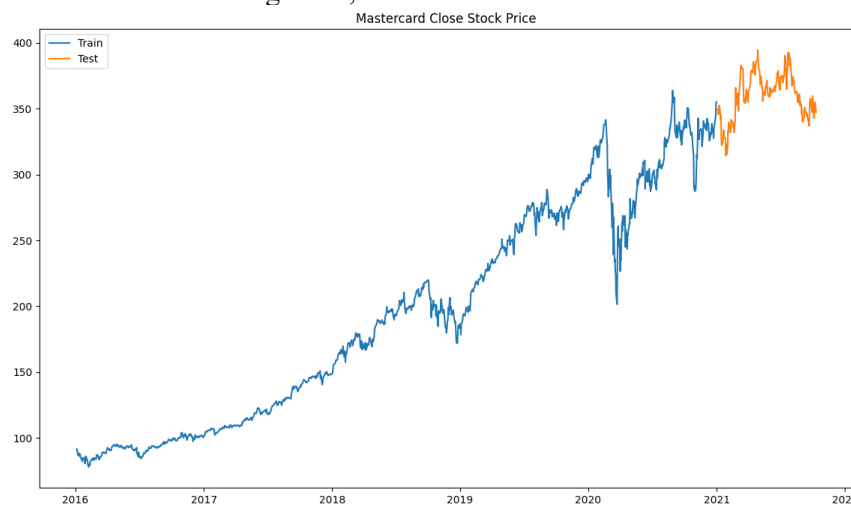

**Figure 3.** Data segmentation.
（Photo credit : Original）

The variation of MAE is recorded during the model training process and the variation curve of MAE in the training set is plotted, and the results are shown in Fig. 4.

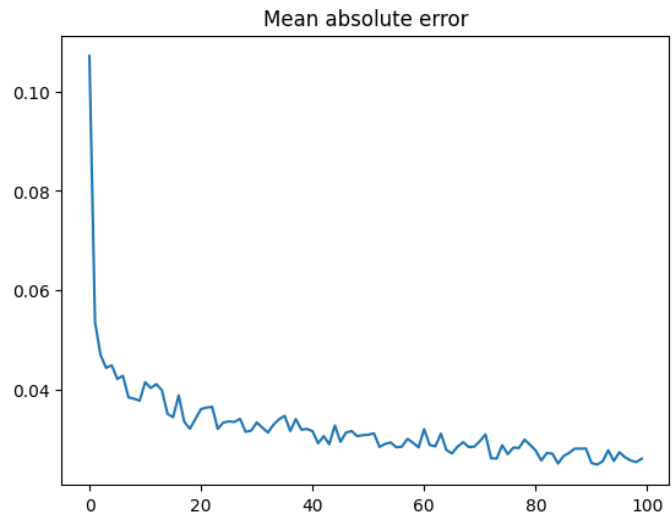

**Figure 4.** MAE.

（Photo credit : Original）

From the change curve of MAE, it can be seen that MAE changes from the initial 0.11 to the final 0.01 and tends to converge, which indicates that the prediction effect of the model is gradually close to the real value, and the model is able to predict the stock price very well.

Use the test set to test the prediction effect of the model and output the line graph between the values of the test set and the predicted values of the model, and the results are shown in Fig. 5.

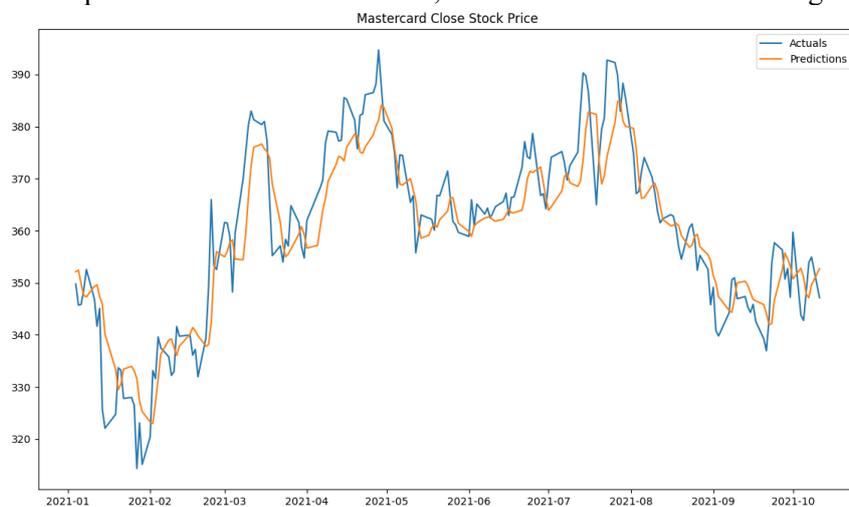

**Figure 5.** Test set prediction results.

（Photo credit : Original）

From the results of stock price prediction by the model in the test set, it can be seen that the time series algorithm based on Genetic Algorithm (GA)-Long Short-Term Memory (LSTM) optimization in this paper is able to predict the stock price very well, which is very close to the actual stock price in terms of trend and value, and the model has a strong generalization ability. The MAE, MSE, RMSE and $R2$ of the test set are also output and the results are shown in Table 3.

**Table 3.** The MAE, MSE, RMSE and R2.

| Evaluation parameters | Value |
| --- | --- |
| MAE | 2.41 |
| MSE | 9.84 |
| RMSE | 3.13 |
| R2 | 0.87 |

## 5. Conclusion

This study provides an innovative approach to stock market forecasting by using a time series algorithm based on optimization of genetic algorithm (GA) and long short-term memory network (LSTM) for stock price prediction. Firstly, after descriptive statistical analysis of the stock data, we introduced an optimized model combining GA and LSTM, and divided the dataset into a training set and a test set for model training and evaluation.

By observing the MAE change curve, we find that the MAE of the model gradually decreases from the initial 0.11 to the final 0.01 and tends to be stable, indicating that the prediction effect of the model is gradually close to the real value and has good convergence. On the test set, we find that the time series algorithm based on GA-LSTM optimization is able to effectively predict the stock price, which is very close to the actual stock price in terms of trend and value, showing good generalization ability. Specifically, the MAE on the test set is 2.41, MSE is 9.84, RMSE is 3.13, and R2 is 0.87.

Taken together, these results show that the GA-LSTM optimization-based time series algorithm proposed in this paper performs well in the stock price prediction task with high accuracy and reliability. This approach not only improves the prediction accuracy, but also demonstrates the advantages of combining the global search of genetic algorithm and the time series modelling capability of LSTM. Therefore, in the context of the current era of computer technology and big data, this novel approach is of great significance for improving stock market prediction and provides useful insights for the financial field.